\documentclass{appolb}
\usepackage{graphicx}
\begin{document}
\title{Deuteron-deuteron collision at 160 MeV%
   \thanks{Presented at Jagiellonian Symposium of Fundamental
and Applied Subatomic Physics, 7-12 June 2015, Krak\'{o}w, POLAND}%
}
\author{G.~Khatri$^{1}$, W.~Parol$^{1}$,  K.~Bodek$^{1}$, I.~Ciepa\l$^{3}$, B. Jamroz$^{2}$ N.~Kalantar-Nayestanaki$^{4}$, St.~Kistryn$^{1}$, B.~K\l os$^{2}$, A.~Kozela$^{3}$, P.~Kulessa$^{3}$, A.~Magiera$^{1}$, I.~Mazumdar$^{5}$, J.~G.~Messchendorp$^{4}$,
D.~Rozp\c{e}dzik$^{1}$, I.~Skwira-Chalot$^{6}$, E.~Stephan$^{2}$, A. Wilczek$^{2}$, B. W\l och$^{7}$, A.~Wro\'{n}ska$^{1}$, J.~Zejma$^{1}$
\address{
  $^{1}$~Jagiellonian University, 30-059 Krak\'{o}w, POLAND\\
  $^{2}$~University of Silesia, 40-007 Katowice, POLAND\\
  $^{3}$~Institute of Nuclear Physics PAN, 31-342 Krak\'{o}w, POLAND\\
  $^{4}$~KVI-CART, University of Groningen, Groningen, The Netherlands\\
  $^{5}$~Tata Institute of Fundamental Research, Mumbai 400 005, INDIA\\
  $^{6}$~Faculty of Physics, University of Warsaw, Warsaw, POLAND\\
  $^{7}$~AGH University of Science and Technology, Krak\'{o}w, POLAND}
}
\maketitle
\begin{abstract}
The experiment was carried out using BINA detector at KVI in Groningen. For the first time an extensive data analysis of the data collected in back part of the detector is presented, where a clusterization method is utilized for angular and energy information. We also present differential cross-sections for the (dd$\rightarrow$dpn) breakup reaction within \textit{dp} quasi-free scattering limit and their comparison with first calculations based on Single Scattering Approximation (SSA) approach. 
\end{abstract}
\PACS{PACS: 21.30.-x, 24.70.+s, 25.10,+s, 13.75.Cs}
  
\section{Introduction}
Experimental studies of three-nucleon dynamics has been the focus of few-body research in recent decades. Among them the nucleon-dueteron (Nd) scattering has been widely investigated. Experiments at KVI in Groningen, at KUTL/RIKEN/RCNP in Japan, and at IUCF in USA have provided large sets of high-precision data \cite{nass} (and references therein), not only for the cross-sections but also for the polarization observables. Tremendous progress has been made to understand the 3N dynamics. With the new high-precision data, covering a large phase-space, it has become possible to pin-down the effects as subtle as three-nucleon forces (3NF) \cite{glock, epel}. The experimental program at KVI has been carried out to extent those studies to the breakup reaction. It used initially the SALAD detector \cite{salad} later upgraded to the BINA detector setup \cite{bina}, covering even larger phase-space and with better detection capabilities. The experiments with SALAD and BINA alone filled up a large gap in the 3N data-base, not only Nd elastic scattering but also breakup reaction~\cite{nass, kistryn2013}. The next step for the experimental program was to move forward in the sector of four-nucleon (4N) system, where the knowledge is scarce in both the theoretical as well as the experimental domains \cite{deltu2}. The 3NF effect is expected to be enhanced in 4N system, that makes the study of 4N even more attractive. 
%%%%%%%%%%%%%%%%%%%%%%%%%%%%%%%%%%%%%%%%%%%%
\section{Experiment}
The BINA detector is a 4$\pi$ apparatus designed for few-nucleon scattering 
experiments at intermediate energies. BINA is divided into two main parts, 
forward wall ($\theta$: 13$^\circ$- 40$^\circ$) and backward ball 
($\theta$: 40$^\circ$- 165$^\circ$). The forward wall consists of (a) multi-wire 
proportional chamber (for reconstruction of angles of the scattered charged 
particles), (b) 24 vertical thin plastic scintillator \lq stripes\rq, and (c) 10 
horizontal thick plastic scintillator \lq slabs\rq. The plastic stripes and slabs 
form 240 $\Delta$E-E telescopes for particle identification. The backward ball 
is nearly spherically symmetric, and made up of 149 triangular phoswich 
detector elements. These elements are arranged in such a way that the formed geometry of the ball resembles to that of a soccer ball. The ball at the same time plays the role of the reaction chamber
 as well as the detector. The work presented here is based on the experiment performed at KVI in Groningen, the Netherlands, with BINA detector, where an unpolarized beam of deuterons with an energy of 160 MeV was provided from AGOR cyclotron to impinge upon liquid hydrogen and liquid deuterium targets.
\section{Data Analysis and Sample Results}
The most basic data analysis steps, in forward scattering (wall) region of BINA, such as particle identification, energy calibration and cross-section evaluation were already described in our previous work~\cite{khatri_fb, khatri_epj, parol_epj}. For the data collected in the backward scattering (ball) region, the analysis task was difficult and challenging. This was mainly due to the lack of light tightness of the ball elements, which resulted in additional contributions from the neighboring elements to the registered signal, i.e. a particle was registered in the ball with more than one element responding. Contribution of such events was significant and could not be neglected.
%----------cluster angle
\begin{figure}[!t]
\centerline{\includegraphics[width=0.7\textwidth]{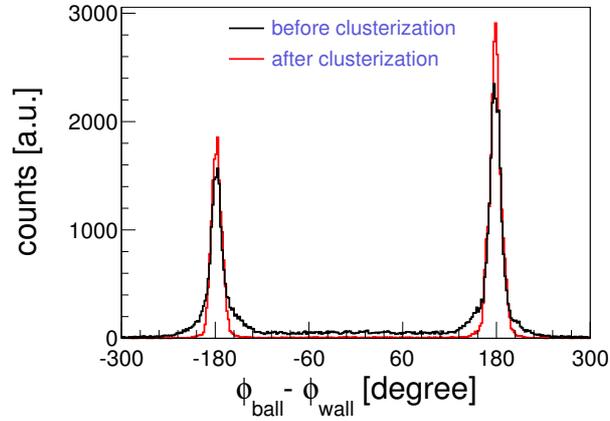}}
\caption[]{The distribution of elastically scattered wall-ball coincident deuterons is presented as a function of their relative azimuthan angle. The peaks are centered around $\pm$180$^{\circ}$ and the width of the peaks represents the angular resolution of the detector. Clusterization (the gray/red curve) led to an improvement in the angular resolution as compared to analysis performed without clusterization (the black curve).}
\label{cluster_angle}
\end{figure}%
%------------- 
%------------wall-ball th-th figure-----wb_theta
\begin{figure}[!b]
\centering
\begin{minipage}{.5\textwidth}
  \centering
  \includegraphics[width=0.98\linewidth]{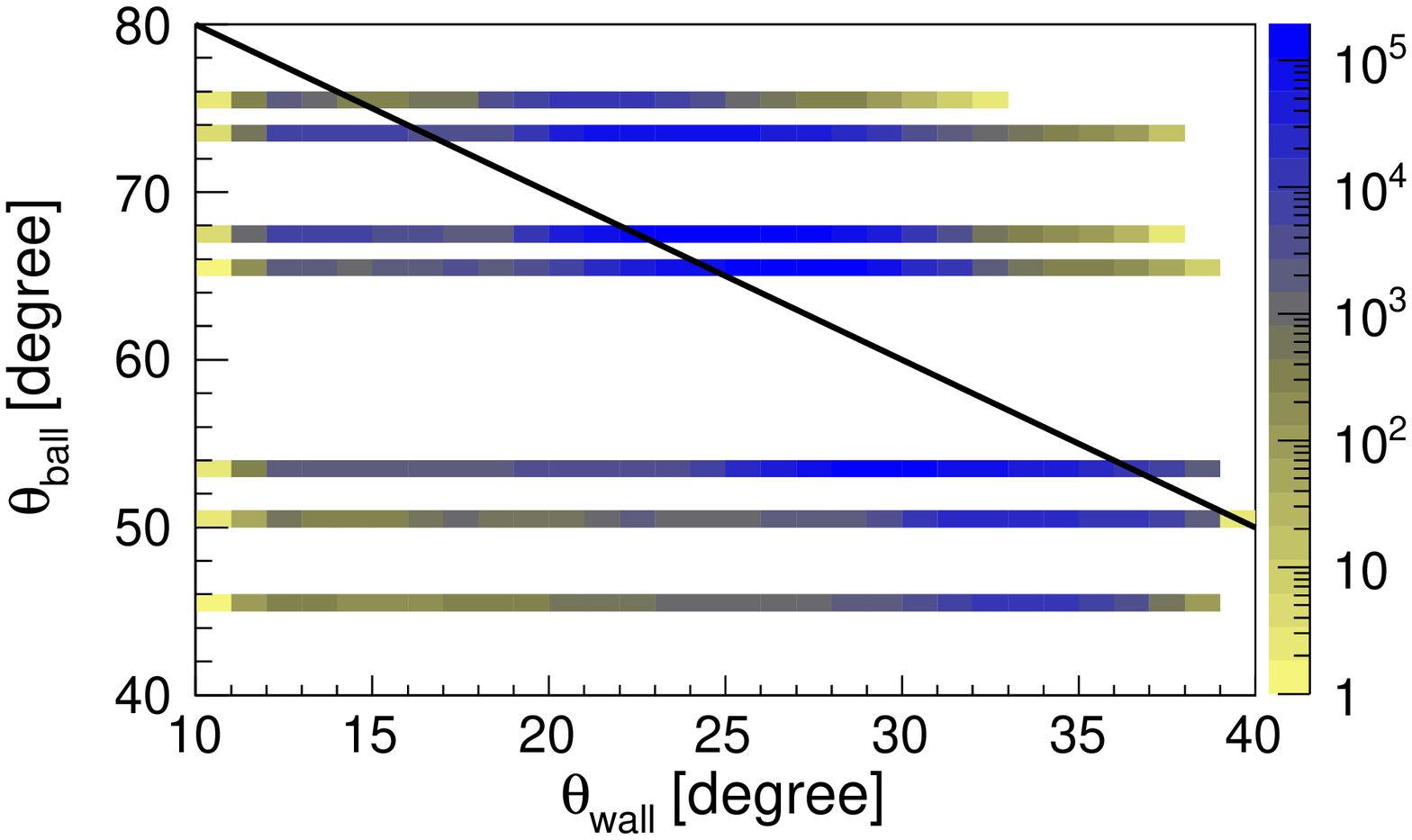}
\end{minipage}%
\begin{minipage}{.5\textwidth}
  \centering
  \includegraphics[width=0.98\linewidth]{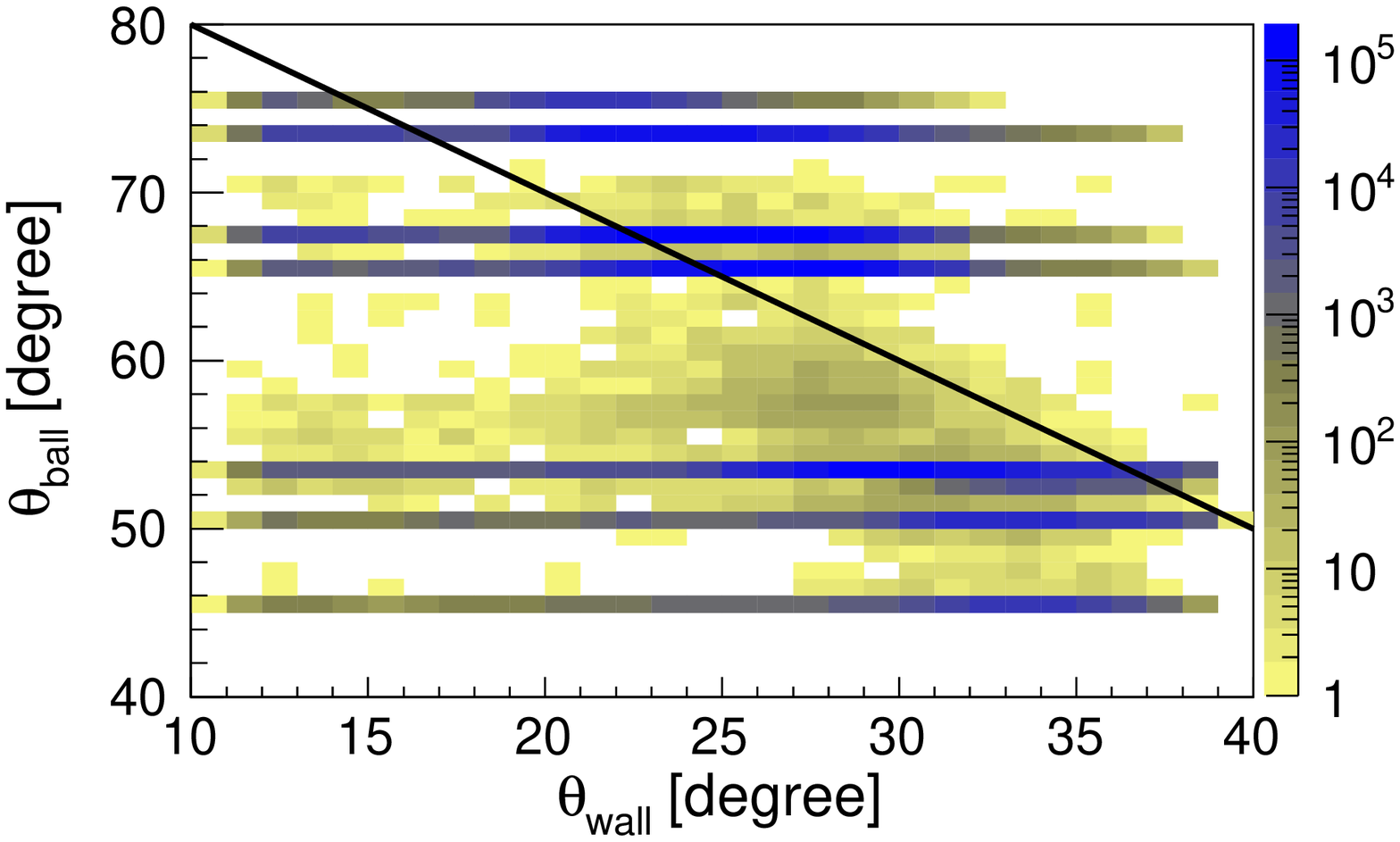}
\end{minipage}
\caption{A correlation between polar angles of wall ($\theta_{wall}$) and ball ($\theta_{ball}$) is presented for the wall-ball coincidence of the two elastically scattered deuterons. Left panel represents before applying the clusterization and the right after. 
}\label{wb_theta_clust}
\end{figure}
%---------------------
Therefore to reconstruct a particle emission angle and energy, a cluster instead of a single element was considered in the further analysis. In an ideal situation, where the ball elements would have been perfectly light-tight, the emission angles ($\theta$, $\phi$) of a detected charged particle in the i$^{th}$ ball element would be taken at the centroid of that ball element, i.e. $\theta$ = $\theta_{i}$  and $\phi$ = $\phi_{i}$ --- these are the angles before applying clusterization. A given cluster is characterized with its azimuthal $\phi_{c}$ and polar $\theta_{c}$ angles and its energy $E_{c}$. The $\phi_{c}$ and $\theta_{c}$ are calculated as weighted average of the angles of the cluster elements as follows:
\begin{equation}\label{clust_eqn1}
\phi_{c}=\frac{\sum\limits_{i=1}^n \phi_{i}E_{i}}{\sum\limits_{i=1}^n E_{i}}\hspace{1cm} and,\hspace{1cm} 
%\end{equation}
%\begin{equation}\label{clust_eqn2}
\theta_{c}=\frac{\sum\limits_{i=1}^n \theta_{i}E_{i}}{\sum\limits_{i=1}^n E_{i}}
\end{equation}
where $n$ is the number of elements constituting a cluster and $i$ refers to the $i^{th}$ element in the cluster. The comparisons of the obtained emission angles, before and after the clusterization, are presented in Fig.~\ref{cluster_angle} (azimuthal) and Fig.~\ref{wb_theta_clust} (polar). The cluster method gives more realistic angular distribution for most of the events, filling the empty gaps.

In order to reconstruct the cluster energy, one needs to take into account a so-called attenuation factor $\alpha$ which refers to the light loss on the borders of the ball elements. Thus the cluster energy is calculated as follows:
%-------------
\begin{figure}[!b]
\centerline{\includegraphics[width=0.6\textwidth]{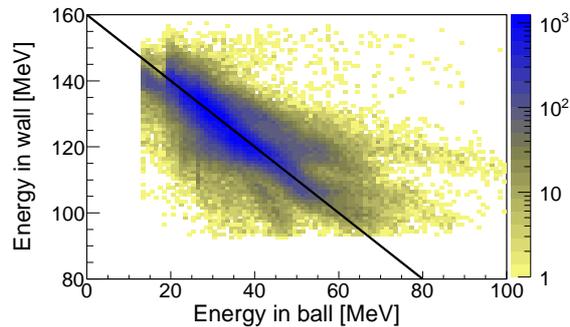}}
\caption[]{The wall-ball energy correlation for the \textit{dd} elastic scattering reaction. The black line refers to the calculated kinematics.}
\label{dp_dd}
\end{figure}%  
%--------------------
\begin{equation}\label{clust_eqn3}
 E_{c} = E_{max} + \sum\limits_{i=1}^{n-1} (1+\alpha) E_{i},
\end{equation}
where $n$ and $i$ have the same meaning as in the Eq.~\ref{clust_eqn1} and the $E_{max}$ is the energy deposited in a central cluster element where the particle is detected (deposits the largest part of its energy). The estimation of $\alpha$ was done by looking at cluster events where only two adjacent ball elements responded to the detection of an elastically scattered deuteron, since the deuteron energy was well known (on the basis of the second deuteron angle measured in wall). Investigation performed for several sample elements showed that the $\alpha$ value is approximately 10\%. The energy calibration, based on clusterization, was checked for the wall-ball coincident dd elastic scattering, see Fig.~\ref{dp_dd}. Further details of the ball detector and the related data analysis can be found in~\cite{khatri_phd}.
%---------sample results--------
\begin{figure}[!t]
\centering
\begin{minipage}{.35\textwidth}
  \centering
  \includegraphics[width=.95\linewidth]{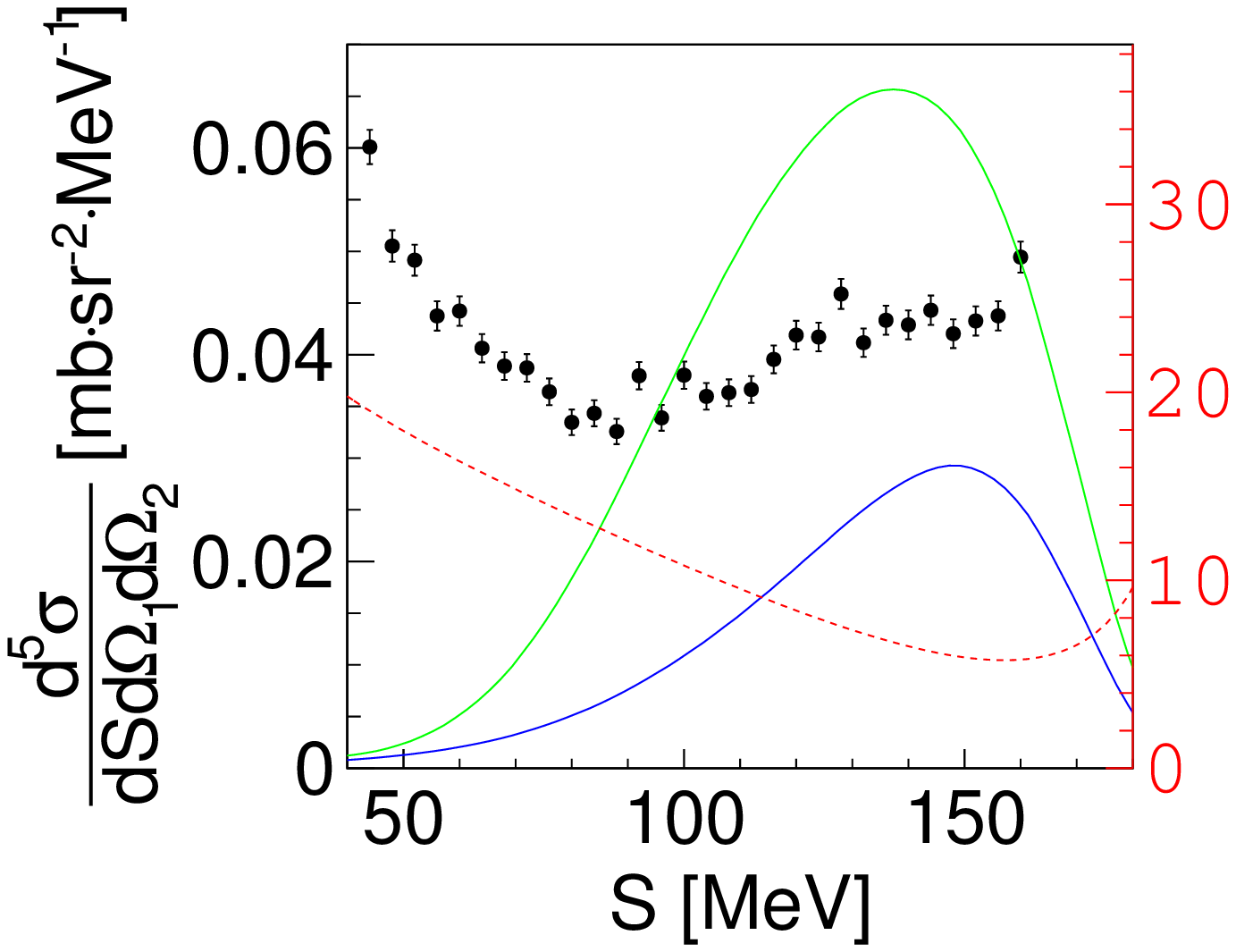}
\end{minipage}%
\hspace{0.1cm}
\begin{minipage}{.3\textwidth}
  \centering
  \includegraphics[width=.95\linewidth]{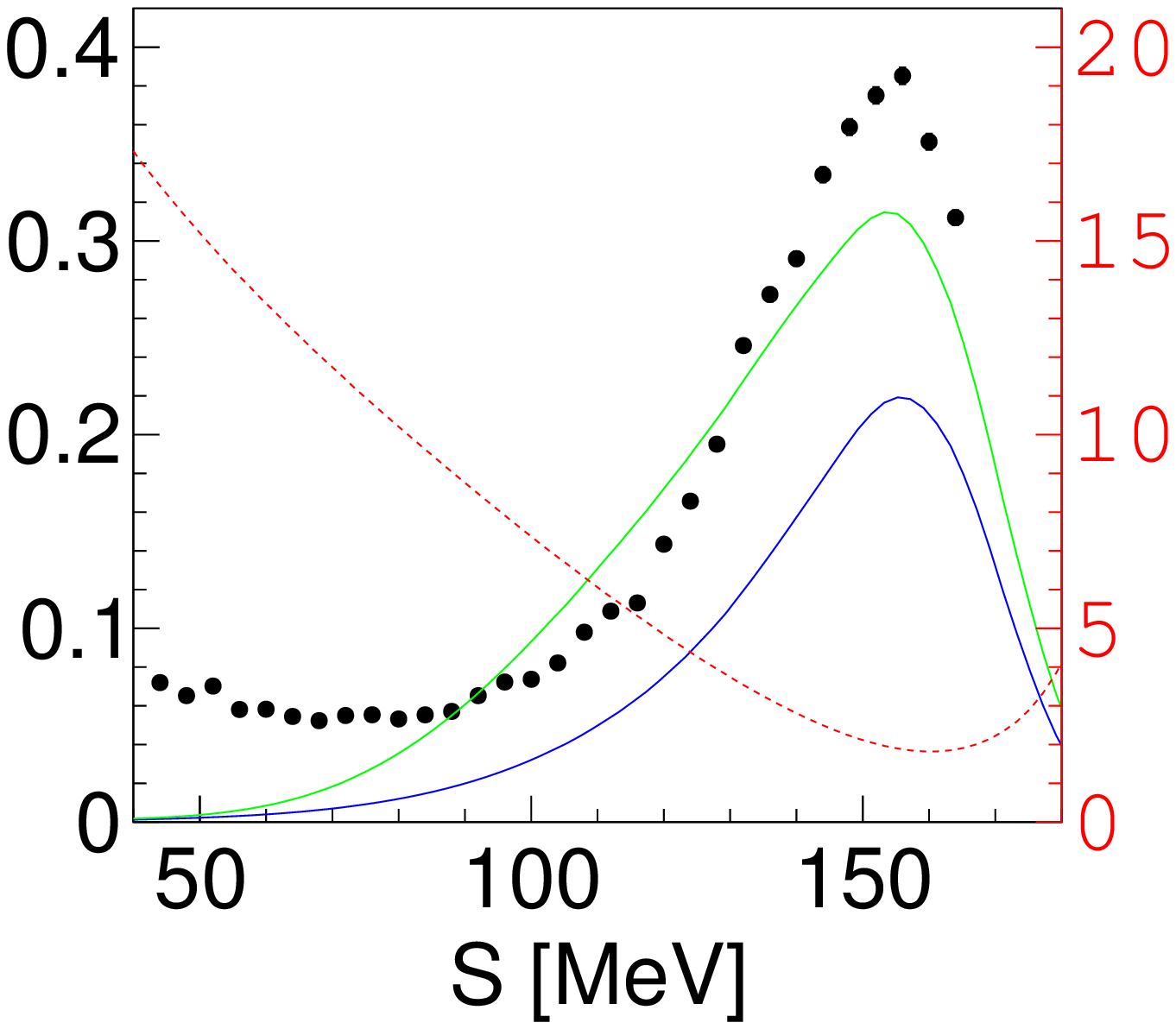}
\end{minipage}
\begin{minipage}{.32\textwidth}
  \centering
  \includegraphics[width=.95\linewidth]{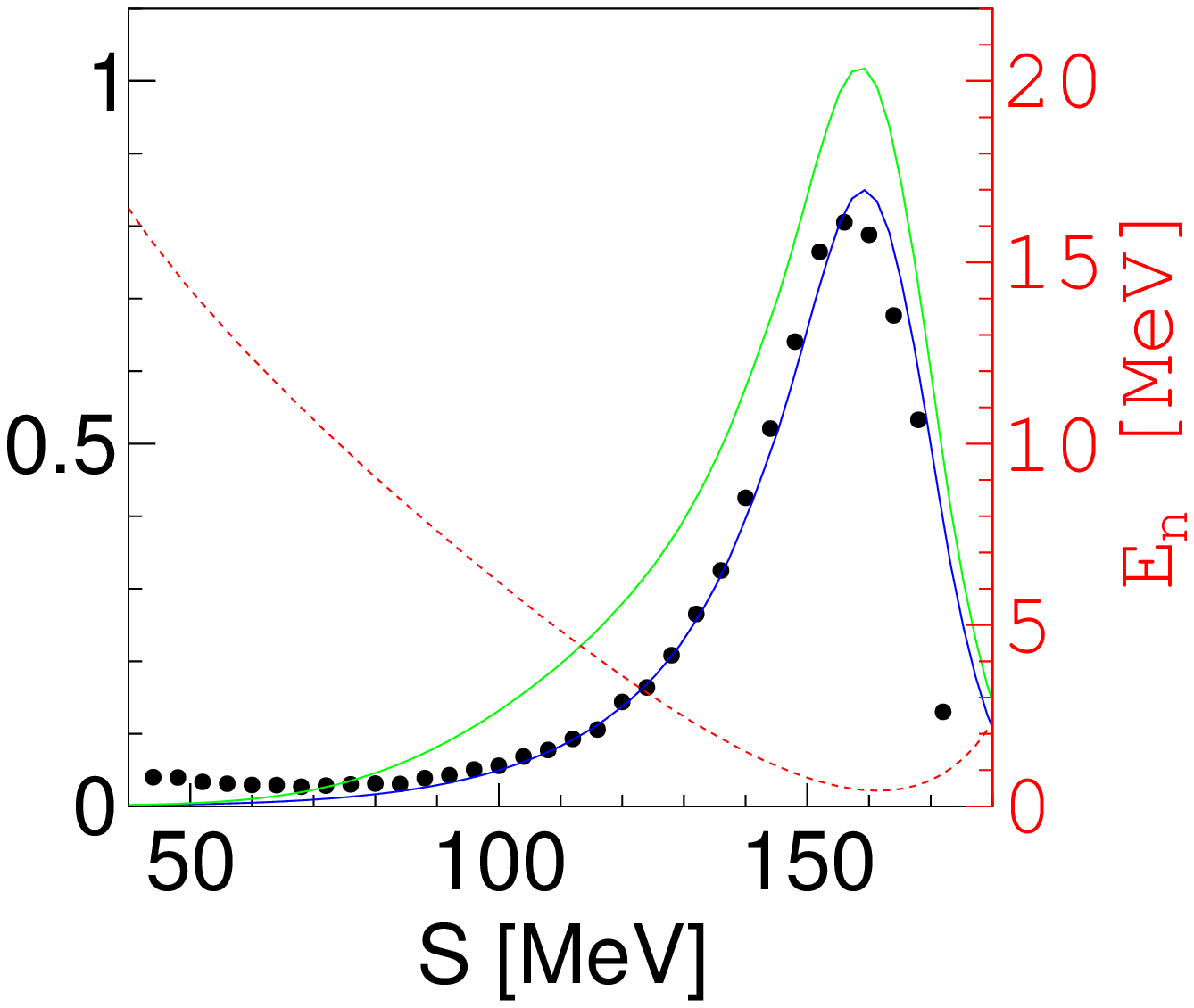}
\end{minipage}
\caption{Sample cross section distributions obtained for 3 geometries characterized by the same combination of polar angles:($\theta_{d}$=22$^{\circ}$, $\theta_{p}$=20$^{\circ}$) and three different $\phi_{dp}$ values: 140$^{\circ}$ (left), 160$^{\circ}$ (center) and 180$^{\circ}$ (right). The solid lines are for theoretical prediction --- blue with 1-term calculations and green with 4-term calculations. The dashed-line and the right hand scale (both in red color) present the dependence of the spectator neutron energy (E$_{n}$) along S-axis. 
}\label{sample_results}
\end{figure}
%-------------
Due to not high enough efficiency of ball, those data were used for checks of systematic effects only. 

For the data collected in the forward wall region of BINA, in the first step the un-normalized differential cross sections for the dd$\rightarrow$dpn breakup reactions were obtained, as it was presented in Ref.~\cite{khatri_epj}. They were subsequently normalized with the use of existing \textit{dd} elastic scattering data from BBS experiment~\cite{bailey}. So far we have obtained the cross sections for 147 kinematical configurations (about 4500 data points) in the dd$\rightarrow$dpn breakup reaction near to quasi-free scattering limit (neutron acting as a spectator). The data, when compared to the state-of-the-art calculations based on Single Scattering Approximation (SSA)~\cite{deltuva_private_comun}, are well described when the neutron energy is close to zero. We present here only sample cross section distributions, see Fig.~\ref{sample_results}. The normalization procedure and the full set of obtained cross section results can be found in ~\cite{khatri_phd}. 
\section{Conclusion and outlook}
We presented a first attempt to precise data analysis of events registered in the ball part of the BINA detector. The obtained cross sections in QFS limit are fairly well described when the spectator neutron energy is small enough. The precise experimental data, so obtained, in a wide phase-space region, can serve as valid tool for verification of rigorous theoretical calculations which have been and are being developed.\\
%uncomment the following lines to place a figure
%\begin{figure}[htb]
%\centerline{%
%\includegraphics[width=12.5cm]{Fig1}}
%\caption{Plot of ...}
%\label{Fig:F2H}
%\end{figure}

{\bf{Acknowledgments}}\\
The author would likes to thank Dr. A. Deltuva for providing the theoretical calculations. We also thank Prof. E. Stephenson and Dr. C. Bailey for allowing us to use their BBS results on elastic scattering for normalization purpose. We acknowledge support by, the Foundation for Polish Science - MPD program, 
co-financed by the European Union within the European Regional Development 
Fund, the "Doctus" stipend by the Lesser Poland region for the PhD students. This work was also supported by the Polish 2013-2016 science found as research Project 2012/05/E/ST2/02313, and by the European Commission within the Seventh Framework Programme through IA-ENSAR (contract no. RII3-CT-2010-262010).

\end{document}